# Exploring the Causal Relationship between Walkability and Affective Walking Experience: Evidence from 7 Major Tertiary Education Campuses in China

**Bojing Liao[1], Jie Zhu[2*]**

[1] Institute of Creativity and Innovation, Xiamen University, 363105 Xiamen, P.R. China

[2] The Data Science Institute, University of Technology Sydney, Australia

[*] Corresponding author

## Abstract

This study investigates the causal relationship between campus walkability and the emotional walking experiences of students, with a focus on their mental well-being. Using data from 697 participants across seven Chinese tertiary education campuses, the study employs a counterfactual analysis to estimate the impact of campus walkability on students' walking experiences. The analysis reveals that students living in campuses with improved walkability are 9.75% more likely to have positive walking experiences compared to those without walkability renovations. While walking attitude is strongly correlated with walking experiences, the study emphasizes the significance of objective factors such as campus surroundings and the availability of walking spaces in influencing the walking experience. Geographical features, including campus walkability improvements, have the most substantial impact, and this effect varies across different subsets of respondents. These findings underscore the importance of considering specific subsets and geographical features when assessing the impact of walkability improvements on the walking experience. In conclusion, the study provides compelling evidence of a causal link between improved campus walkability and enhanced emotional walking experiences among students, suggesting the need for further research on mediating factors and cultural variations affecting student mental health on various Chinese campuses.

**Keywords:** Campus Walkability; Walking experience; Causal inference; Counterfactual Analysis

# Introduction

Walkability, a key evaluation tool, measures the suitability of a built environment for walking, whether for physical activity, active mobility, leisure, or amenity access [1–3], and contributes to social interaction and community engagement [4,5]. Several empirical investigations have identified crucial urban-space constituents for walkability analysis, such as density, diversity, design, destination, distance, and other environmental factors [4,6–12]. A pedestrian-oriented environment reduces motorized vehicle dependence, conserves nonrenewable resources, and minimizes neighborhood pollution [2,13–15], leading to its recommendation in the Global Action Plan on Physical Activity 2018-2030 [16].

One of the key functions of university campuses is to promote student health and fitness through walkability [3,17–20]. The walkability is typically determined by the built environment factors and student perceptions [17,19,21,22]. Compare to western campuses, Chinese campus has defined boundaries, guarded entrances, and universally accessible amenities [17,19], which aims fostering healthy living habits and personal growth [3,23,24]. Previous research indicates that the walkability is associated with social capital, identity, engagement, safety and livability within different neighborhoods [25–30]. These factors, in turn, have been found to significant contribute to individuals' affective walking experience (i.e., hedonic and eudemonic elements) and mental health [31,32].

However, the causal relationship between campus walkability and students' affective walking experiences remains insufficiently examined. Previous research into the non-causal relationship has produced inconsistent results, such as Lee and Shepley (2020), who reports positive influences of campus factors on walking experience [33], and Keat et al. (2016), who do not [34]. These discrepancies may reflect regional differences and a lack of comprehensive evidence regarding the impact of various walkable characteristics. The role of walking attitude, a key factor relates to perceived walkability and subsequent walking behaviors and experiences [35–37], is also significant. However, in a campus context, it is unclear to what extent walking attitude influences the affective walking experience when walkability is accounted for.

This article attempts to estimate the causal impact of campus walkability on affective walking experience, accounting for individual walking attitudes. Our aim is to analyze the causal effects of campus walkability on the affective walking experience according to compare with different campuses (campus with or without renovation). To achieve this, we conduct a campus environmental survey to evaluate campus walkability, individual walking attitude, and affective walking experience. We employ counterfactual analysis to estimate the causal effect of campus walkability on affective walking experience. Overall, utilizing these methodologies to comprehend campus walkability and its causal effect on affective walking experience could ameliorate not only the environment but also perceptions of campus members.

# Methods

## Study Design and Population

The research is carried out in Wuhan City, a significant urban hub in the central-southern region of China, encompassing seven university campuses (refer to Table 1). The campuses that underwent renovations are Huazhong University of Science and Technology, Wuhan University, and China University of Geosciences (Wuhan), while those without any refurbishment are Central China Normal University, Wuhan University of Technology, Zhongnan University of Economics and Law, and Huazhong Agricultural University. All of these institutions are directly affiliated with the Chinese Ministry of Education and are classified as Class A Double First-Class Universities. Each campus possesses distinct features, including access gates, well-defined boundaries, independent services and facilities, and large dormitory areas. These characteristics contribute to their resemblance to local neighborhoods. More importantly, these campuses are exclusively for students and faculty, which is why we chose these areas for the study.

An online survey conducts to collect data and analyze the causal relationship between campus walkability, the affective walking experience, and walking attitude. We utilize the Chinese version of the abbreviated Neighborhood Environment Walkability Survey (NEWS-A) to measure campus walkability [38]. Given that every participant in our study resides in a four-person student apartment, we intentionally do not consider the perceived residential density as it is contingent upon the type of housing. Hence, this survey comprised five multi-item subscales and four single-item subscales. Five multi-items are land use mix-diversity, access to services, street connectivity, infrastructure and safety for walking. And, four single items include aesthetics, traffic safety, access to parking, hilly streets, physical/nature obstacles, and presence of dead-end streets [11,39]. With the exception of the land use mix-diversity rating scale, the remaining four subscales are assessed on a scale ranging from 'strongly disagree' (1) to 'strongly agree' (5). The land use mix-diversity is determined by the walking distance from the apartment to various types of shops and amenities, with response options ranging from '1 to 5 minutes' to 'over 30 minutes' [7,11,38,39]. A higher score on this scale indicates closer average proximity.

*Table 1. The basic information of study site, all information comes from the official university website.*

| | Name of institution | Number of Academic staff | Number of students | Number of campus | Campus area in total |
|---|---|---|---|---|---|
| **With renovation** | Huazhong University of Science and Technology | 3448 persons | 56040 persons | 2 | 514 hectares |
| | Wuhan University | 7325 persons | 58720 persons | 1 | 320 hectares |
| | China University of Geosciences (Wuhan) | 3122 persons | 31040 persons | 2 | 177 hectares |
| **Without renovation** | Central China Normal University | 4700 persons | 31600 persons | 1 | 330 hectares |
| | Wuhan University of Technology | 5570 persons | 52000 persons | 3 | 174 hectares |
| | Zhongnan University of Economics and Law | 1173 persons | 26400 persons | 2 | 206 hectares |
| | Huazhong Agricultural University | 1443 persons | 24765 persons | 1 | 495 hectares |

In addition, we employ scale items from Cao et al. (2006) to measure walking attitude, specifically: 'I like walking' and 'If possible, I would rather walk than drive' [40]. The scores for these two items range from 'strongly dislike' (1) to 'strongly like' (5). For the affective walking experience, empirical research has indicated that happiness, comfort, annoyance, and security are correlated with perceived walkability (walking through experience) [41–44]. Therefore, we use statements such as 'I felt happy/ comfortable/ annoyed/ secure' to assess the affective walking experience. Respondents rated each affective statement on a 5-point Likert scale, ranging from 'strongly disagree' (1) to 'strongly agree' (5). In the survey, participants are also asked to provide their gender, level of education, duration of campus residency, cultural background (native or not), and daily walking frequency.

## Data Analysis

Data for this study is collected from April to July 2021. The respondents, all students from seven institutions, complete the questionnaire via a nationwide online research platform. We recruit participants evenly from twelve campuses across these institutions, with a total of 1026 students completing the questionnaire. To ensure data quality, we exclude respondents who either provided identical answers to all questions or completed the NEWS-A section in less than 180 seconds. After the data cleaning process, the final sample include 697 students. Figure 1 and Table 2 describe the sample characteristics of campuses, distinguishing between those that have undergone renovations and those that have not.

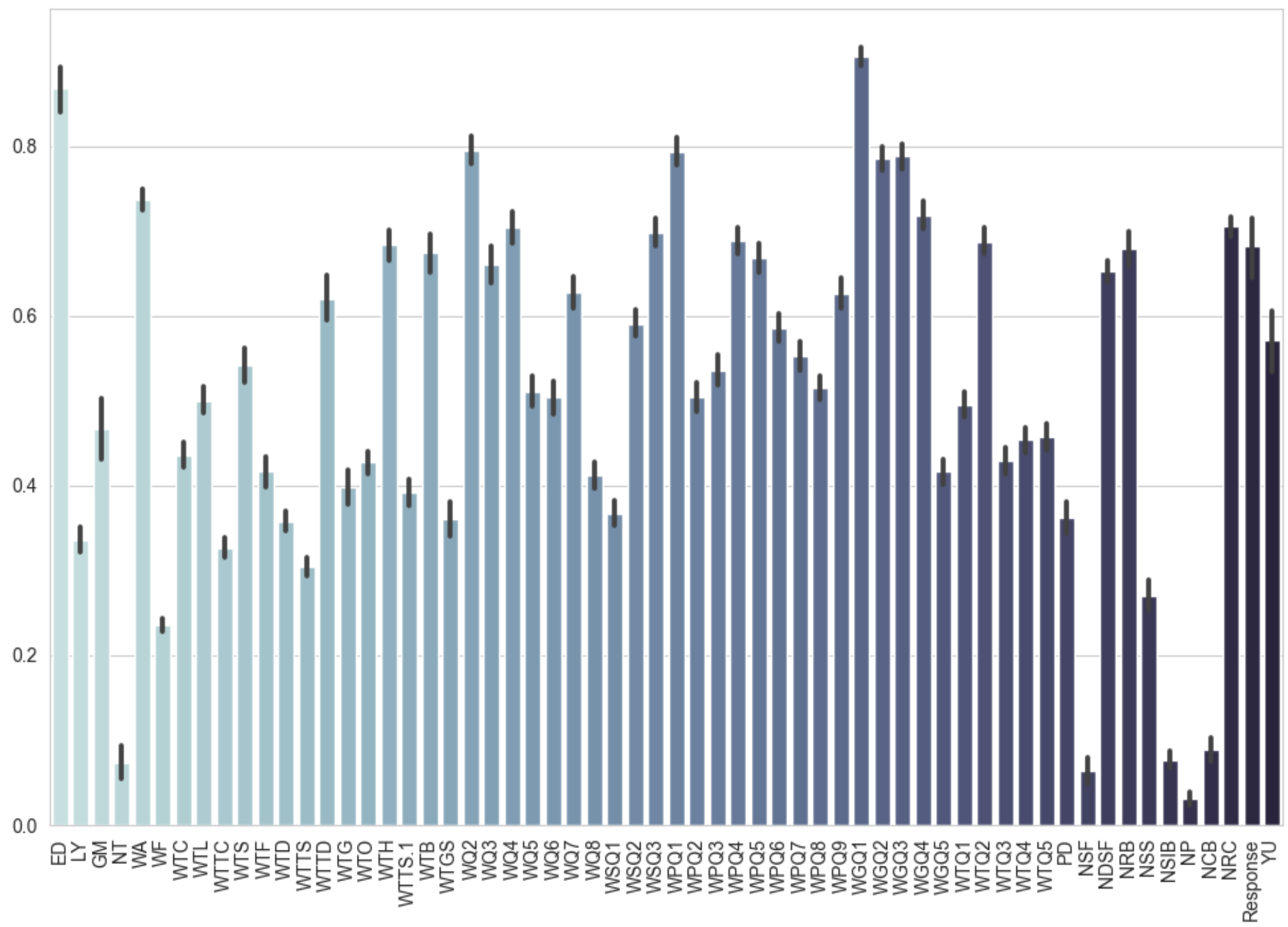

*Figure 1. The average value for each question in the survey, the response variable and intervention (YU), computed across all respondents. The horizontal axis represents the survey questions, while the vertical axis represents the average response value. The error bars indicate the 95% confidence interval of the mean. The figure provides an overview of the respondents' overall attitudes and experiences related to walking on campus, as captured by the survey questions.*

According to empirical research, perceived walkability is also related to geographical features such as density and the number of amenities [7,11,17,28,29,38]. We hence incorporate additional geographical features generated by ArcGIS to the survey data. Table 3 provides descriptive analysis information and abbreviations for all measurement items in the dataset.

*Table 2. The sample characteristics description.*

| | Abbreviation | Sample (N) | Sample (%) |
|---|---|---|---|
| ***Gender*** | | | |
| Male students (with renovation) | | 164 | 23.5 % |
| Male students (without renovation) | GM | 162 | 23.2% |
| Female students (with renovation) | | 193 | 27.7 % |
| Female students (without renovation) | | 178 | 25.6% |
| ***Education level*** | | | |
| Undergraduate students (with renovation) | | 282 | 40.5 % |
| Undergraduate students (without renovation) | ED | 323 | 46.3 % |
| Graduate students (with renovation) | | 75 | 10.8 % |
| Graduate students (without renovation) | | 17 | 2.4 % |
| ***Length of campus residence*** | | | |
| Less than one year (with renovation) | | 251 | 36.0 % |
| Less than one year (without renovation) | | 141 | 20.2% |
| 1-2 years (with renovation) | LY | 93 | 13.3 % |
| 1-2 years (without renovation) | | 105 | 15.1 % |
| More than 2 years (with renovation) | | 13 | 1.8 % |
| More than 2 years (without renovation) | | 94 | 13.6% |
| ***Cultural background*** | | | |
| Native students (with renovation) | | 31 | 4.4% |
| Native students (without renovation) | NT | 20 | 2.9% |
| Non-native students (with renovation) | | 326 | 46.8 % |
| Non-native students (without renovation) | | 320 | 45.9% |
| ***Total (with renovation)*** | | 357 | 51.2% |
| ***Total (without renovation)*** | | 340 | 48.8% |
| ***Total (all)*** | | 697 | 100 % |

*Table 3. Descriptive analysis and abbreviations for all measurement items.*

| Variables | Abbreviation | Mean | Sd. |
|---|---|---|---|
| ***Campus walkability*** | | | |
| *1. About how long would it take to get from your dormitory to the nearest facilities listed below if you walked to them* | | | |
| Classroom buildings | WTC | 0.44 | 0.20 |
| Library | WTL | 0.50 | 0.21 |
| Canteen | WTTC | 0.33 | 0.16 |
| Snack bar | WTS | 0.54 | 0.27 |
| Fruits store | WTF | 0.42 | 0.25 |
| Delivery points | WTD | 0.36 | 0.16 |
| Supermarket | WTTS | 0.30 | 0.14 |
| Department store | WTTD | 0.62 | 0.36 |
| Gym | WTG | 0.39 | 0.28 |
| Outdoor sports space | WTO | 0.43 | 0.19 |
| Hospital | WTH | 0.68 | 0.25 |
| School bus stop | WTTS1 | 0.39 | 0.22 |
| Bus stop | WTB | 0.67 | 0.31 |
| Green space | WTGS | 0.36 | 0.28 |
| *2. Access to services* | | | |
| Classroom buildings are within easy walking distance of my dormitory | WQ1 | 0.76 | 0.26 |

| Variables | Abbreviation | Mean | Sd. |
|---|---|---|---|
| Stores are within easy walking distance of my dormitory | WQ2 | 0.79 | 0.22 |
| Commercial areas are easy to arrive by public transportation | WQ3 | 0.66 | 0.28 |
| There are many places to go within easy walking distance of my dormitory | WQ4 | 0.70 | 0.25 |
| There are many slopes that make the road difficult to walk | WQ5 | 0.51 | 0.23 |
| There are many obstacles that make the road difficult to walk | WQ6 | 0.50 | 0.26 |
| There are many pedestrians on the street at peak hours | WQ7 | 0.63 | 0.25 |
| There are many pedestrians on the street at non-peak hours | WQ8 | 0.41 | 0.21 |
| *3. Streets on my campus* | | | |
| The streets on my campus do not have many, or any, cul-de-sacs | WSQ1 | 0.37 | 0.19 |
| The distance between intersections on my campus is usually short | WSQ2 | 0.59 | 0.22 |
| There are many alternative routes for getting from place to place | WSQ3 | 0.69 | 0.22 |
| *4. Places for walking* | | | |
| There are sidewalks on most of the streets on my campus | WPQ1 | 0.79 | 0.22 |
| Sidewalks are occupied by parked cars on my campus | WPQ2 | 0.50 | 0.23 |
| There is a strip that separates the streets from the sidewalks on my campus | WPQ3 | 0.54 | 0.25 |
| Streets are well-lit at night | WPQ4 | 0.69 | 0.23 |
| There are zebra crossings and traffic lights on a busy street | WPQ5 | 0.67 | 0.25 |
| There are many manholes covers in the sidewalks | WPQ6 | 0.59 | 0.21 |
| Walkable indoor spaces (with air-conditioning) are available on my campus | WPQ7 | 0.55 | 0.24 |
| Streets and sidewalks are usually wet on my campus | WPQ8 | 0.52 | 0.20 |
| There are rest facilities, such as benches | WPQ9 | 0.63 | 0.23 |
| *5. Campus surroundings* | | | |
| There are trees along the streets on my campus | WGQ1 | 0.91 | 0.15 |
| There are many interesting things to look at while walking on my campus | WGQ2 | 0.79 | 0.20 |
| There are many green spaces on my campus | WGQ3 | 0.72 | 0.22 |
| There are attractive buildings on my campus | WGQ4 | 0.42 | 0.19 |
| Air pollution is usually high on my campus | WGQ5 | 0.49 | 0.22 |
| *6. Safety from traffic* | | | |
| There is so much traffic along the nearby street that it makes it difficult to walk | WTQ1 | 0.49 | 0.22 |
| The speed of traffic on most nearby streets is usually slow (30 km/h or less) | WTQ2 | 0.69 | 0.22 |
| Most drivers exceed the posted speed limits while driving on my campus | WTQ3 | 0.43 | 0.20 |

| Variables | Abbreviation | Mean | Sd. |
|---|---|---|---|
| There are many parked cars on nearby streets that which makes it difficult to cross | WTQ4 | 0.45 | 0.20 |
| There are many passing cars on nearby streets that it is frightening to cross | WTQ5 | 0.46 | 0.22 |
| ***Geographical features*** | | | |
| Population density (persons/square kilometer) | PD | 0.36 | 0.25 |
| Number of sports facilities within 1 km | NSF | 0.06 | 0.21 |
| Number of daily service facilities within 1 km | NDSF | 0.65 | 0.17 |
| Number of residential buildings within 1 km | NRB | 0.68 | 0.28 |
| Number of supermarkets of shopping mall within 1 km | NSS | 0.27 | 0.24 |
| Number of scientific institutional buildings within 1 km | NSIB | 0.07 | 0.15 |
| Number of parking within 1 km | NP | 0.03 | 0.11 |
| Number of companies or business within 1 km | NCB | 0.09 | 0.19 |
| Number of restaurants or canteen within 1 km | NRC | 0.71 | 0.16 |
| ***Affective walking experience (positive walking emotion)*** | | | |
| I felt happy | WEQ1 | 0.75 | 0.19 |
| I felt annoyed | WEQ2 | 0.44 | 0.50 |
| I felt comfortable | WEQ3 | 0.74 | 0.19 |
| I felt secure | WEQ4 | 0.76 | 0.18 |
| **Walking attitude** | WA | 0.74 | 0.17 |
| **Walking frequency** | WF | 0.24 | 0.10 |

*Notes: important features for variations in treatment effect are shaded in light gray.*

## Average Treatment Effect Estimation

In this study, we employ the counterfactual estimator [45] to estimate the effect of enhanced campus walkability on the affective walking experience of students surveyed across seven prominent tertiary education campuses in China. The counterfactual framework is a non-parametric, data-driven method that enables the identification of heterogeneous treatment effects by recursively partitioning the data into subgroups based on their covariate values. By utilizing this method, we aim to provide a robust and interpretable estimate of the causal effect of improved campus walkability on the affective walking experience of the surveyed students, which may have significant implications for campus planning and management.

To construct the counterfactual estimator, we define the binary treatment variable $\mathrm{A_i}$, with $\mathrm{A_i} = 0$ indicating that individual $i$ belongs to the control group and their campus has not undergone any significant construction works to improve the campus walkability and $A_i = 1$ indicating the treatment group with significant construction works. The outcome of interest is defined based on the responses of the surveyed students to four questions, denoted as $WEQ_1$ , $WEQ_2$ , $WEQ_3$ , $WEQ_4$. Specifically, we define $YU_i$ as the indicator function of a positive response to a combination of three questions, $WEQ_1$ , $WEQ_3$ , $WEQ_4$ minus a negative response to question $WEQ_2$. Mathematically, this can be expressed as

$$YU_i = \mathbb{1}(WEQ_1 + WEQ_3 + WEQ_4 - WEQ_2 > 1),$$

where $\mathbb{1}$ is the indicator function that takes a value of 1 if the evaluated condition is true, and 0 otherwise. This definition of $YU_i$ is chosen to capture the effect of improved campus walkability on the overall affective walking experience of the surveyed students, as perceived through their responses to the four survey questions.

The counterfactual effect of the treatment $A_i$ on the outcome variable of interest, denoted as the Average Treatment Effect (ATE), can be obtained by comparing the expected outcome under treatment to the expected outcome under control, i.e.,

$$\mathrm{ATE_i} = \mathrm{E}[\mathrm{Yu_i}|\mathrm{A_i} = 1] - \mathrm{E}[\mathrm{YU_i}|\mathrm{A_i} = 0].$$

However, since we only observe the outcome for everyone under one of the treatment conditions, we need to estimate the potential outcomes for everyone under both treatment and control conditions, which are unobserved. The algorithm estimates the ATE by recursively partitioning the data into subgroups based on the covariate values, and then comparing the average outcome of the treated group to that of the control group within each subgroup. The resulting treatment effect estimate for each subgroup is then weighted by the proportion of individuals in that subgroup to obtain the overall ATE. The counterfactual effect of the treatment on everyone can also be estimated by applying the algorithm to the individual-level data and comparing the predicted outcome under treatment to the predicted outcome under control. This approach, known as the individual treatment effect (ITE) estimation, allows us to identify the subgroups of individuals who are most likely to benefit from the treatment, and can inform personalized treatment recommendations. To apply the framework, we employ a random forest model that take the survey responses to questions 1 through X as input features and the responses to the Walking Experience Questionnaire (WEQ) items 1 through 4 as the response variables. The treatment variable is defined as whether the campus underwent significant renovation to improve walkability. Our random forest model is trained to estimate the expected outcome for each respondent under both the treated and untreated conditions. By comparing the two expected outcomes, we could estimate the treatment effect of the walkability improvements. This approach allows us to estimate the causal effect of the treatment, while adjusting for potential confounding factors captured by the survey responses.

# Results

The survey data reveals an average response value of 0.682 (the 95% confidence interval is from 0.6485 to 0.7172), which implies a positive walking emotion from students. The unadjusted average difference in responses between individuals residing in an improved campus and those in the old campus is 5.58%.

The counterfactual estimator produced an adjusted difference (or average treatment effect) of 9.75% (8.60%, 10.90%), which is 4.17% higher than the unadjusted difference. This effect indicates that individuals residing in a campus that underwent substantial walkability improvements are 9.75% more likely to report a positive walking experience on campus (Y=1), after adjustments for the recorded survey responses

It is important to note that the estimated treatment effect is obtained using the fitted counterfactual model, which uses recorded survey responses to predict the treatment effect. Thus, the estimated effect may not necessarily generalize to other models or contexts. Furthermore, the range of the estimated effect provides an indication of the level of uncertainty associated with the estimate and should be taken into consideration when interpreting the results.

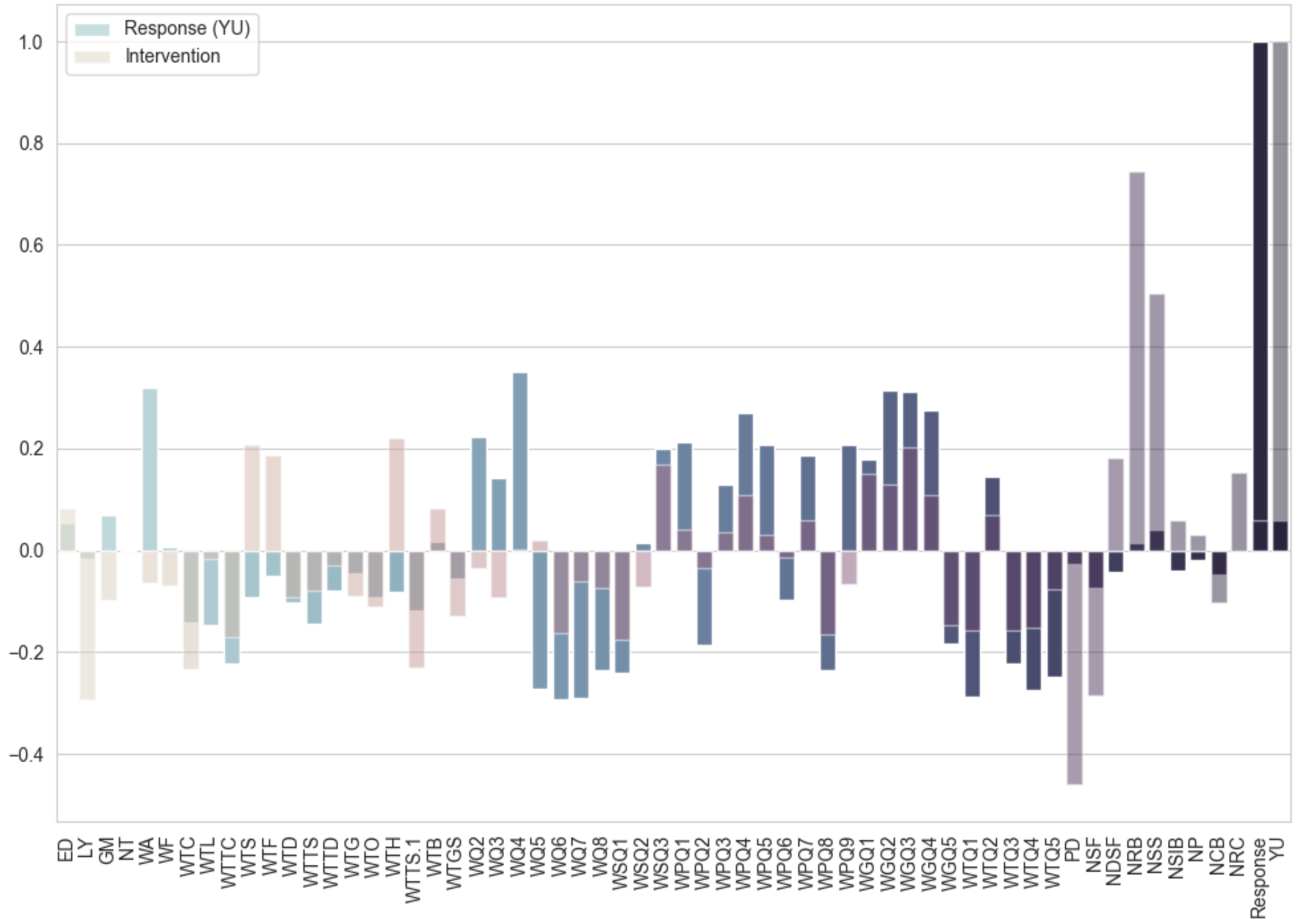

*Figure 2. The correlation among survey questions with the response variable and the intervention variable (YU).*

In contrast to correlation-based analytics, counterfactual analysis provides us with insights into the factors driving the differences in treatment effects. For instance, Figure 2 demonstrates that 'Walking attitude' (WA) and the attitude towards the statement 'there are many places which are easy to access by walking' (WQ4) exhibit the strongest correlation with the response variable, which represents the walking experience. Conversely, the intervention (improvement on campus walkability) shows minimal correlation with the response. A regression model would conclude that the improvement on campus walkability is not a significant driver of the campus walking experience. Instead, it would suggest that factors such as students' current walking attitude, campus surroundings (WGQ) and the availability of places for walking (WPQ) are important, as they exhibit higher correlations with the response.

The counterfactual analysis enables us to delve deeper into the factors that play a pivotal role in driving the variation in treatment effects. This approach provides a more accurate understanding of the key influencers of the walking experience, surpassing the limitations of relying solely on correlation-based methods. By conducting counterfactual analysis, we can effectively account for potential confounding factors, as demonstrates in Figure 2, where features like WGQ 1 to 4 exhibit high correlations with both the response variable and the intervention. This ensures a more rigorous and unbiased examination of the factors impacting the walking experience.

Upon fitting the counterfactual model, we calculate the Mean Decrease Impurity (MDI) values to identify the features that has the highest importance in predicting the treatment effect.

Figure 3 demonstrates that geographical features, specifically NSIB (Number of scientific institutional buildings within 1 km) and NSS (Number of supermarkets or shopping malls within 1 km), exhibit the highest importance for the variance in the difference in campus walking experience. These findings suggest that, in addition to the physical enhancements make to campus facilities (the treatment), geographical factors play a crucial role in influencing the treatment effect of walkability improvements. On the other hand, subjective factors such as walking attitude (WA), campus surroundings (WGQ), and the availability of places for walking (WPQ) have minimal impact on the variation of the treatment effect. While these subjective feelings may still contribute to the overall walking experience, their influence on the treatment effect of walkability improvements appears to be less significant compared to the geographical features.

By considering these adjustments and examining the relative importance of different factors, we gain a better understanding of the specific influences on the treatment effect and can identify the key influencers beyond subjective perceptions.

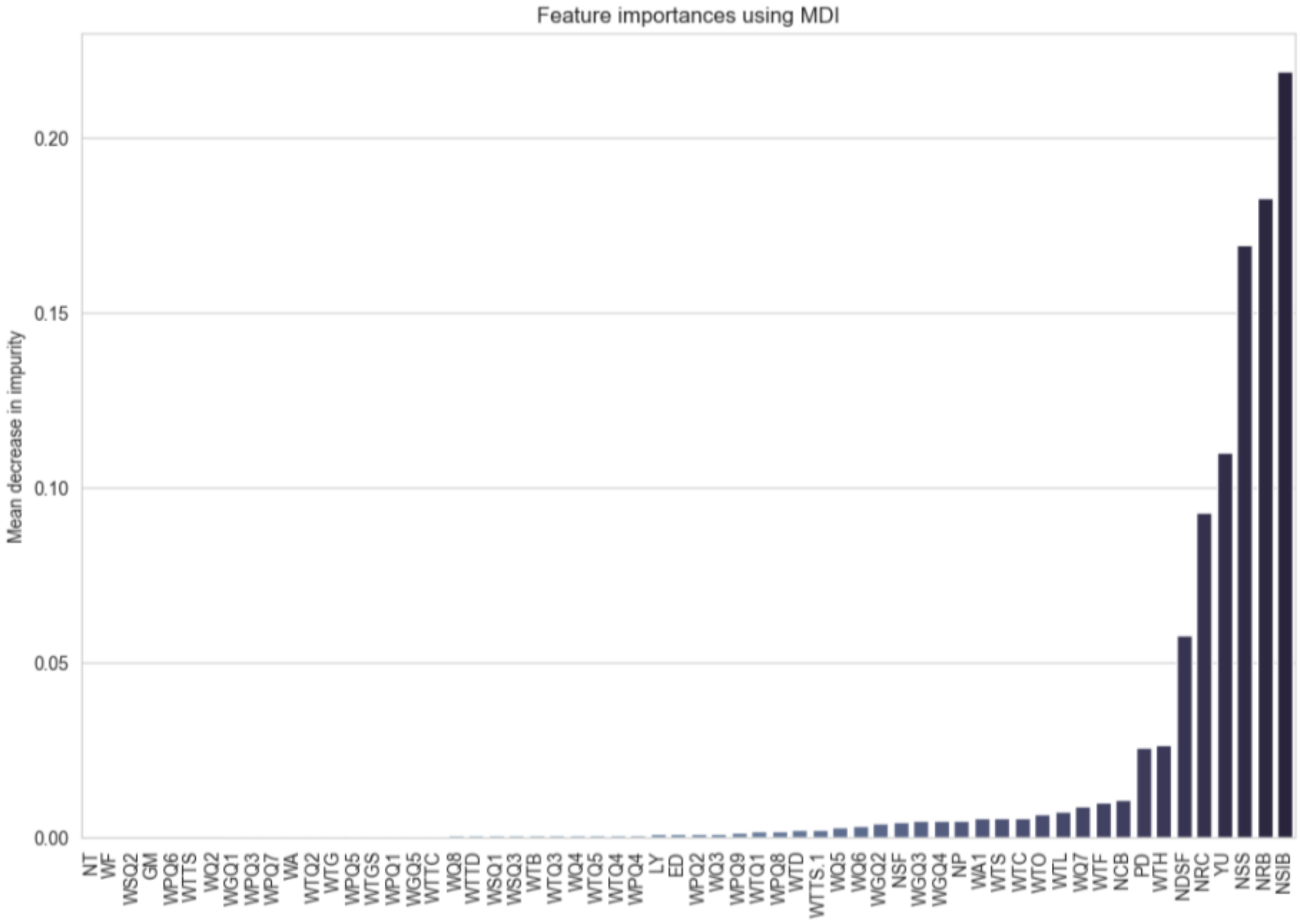

*Figure 3. The feature importance values as computed by the Mean Decrease Impurity (MDI) method using the predicted treated effect as the response variable and all features as recorded in Table 2 and 3 as well as the intervention (YU) as the predictors.*

To investigate how the treatment effect may vary based on NSIB and NSS values, we predict the potential treatment effect across a grid of integer values ranging from 0 to 5 for each feature. This results in a total of 36 distinct combinations. We then enter these values into a data matrix, with all other survey responses set to their average values observed in the original dataset. Figure 4 displays the resulting heat map of predicted treatment effects across the grid of NSIB and NSS values.

Our findings suggest that the maximum treatment effect is observed when NSIB and NSS values reached their highest levels. This implies that individuals whose campuses are situated in areas with a greater number of institutional buildings and shopping malls are more likely to experience a stronger treatment effect. To further investigate, we conduct a similar analysis but condition on the median value of all other survey responses. Surprisingly, we arrive at a similar conclusion, indicating that the geographical factors of NSIB and NSS continue to play a significant role in influencing the treatment effect.

However, when considering the median value, the predicted treatment effect is approximately 10% higher than the one obtained using the average value. This implies that the treatment effect may vary depending on the specific characteristics of the respondents and their campuses. It highlights the importance of considering different subsets of survey respondents and their corresponding values when evaluating the impact of geographical features on the treatment effect of walkability improvements.

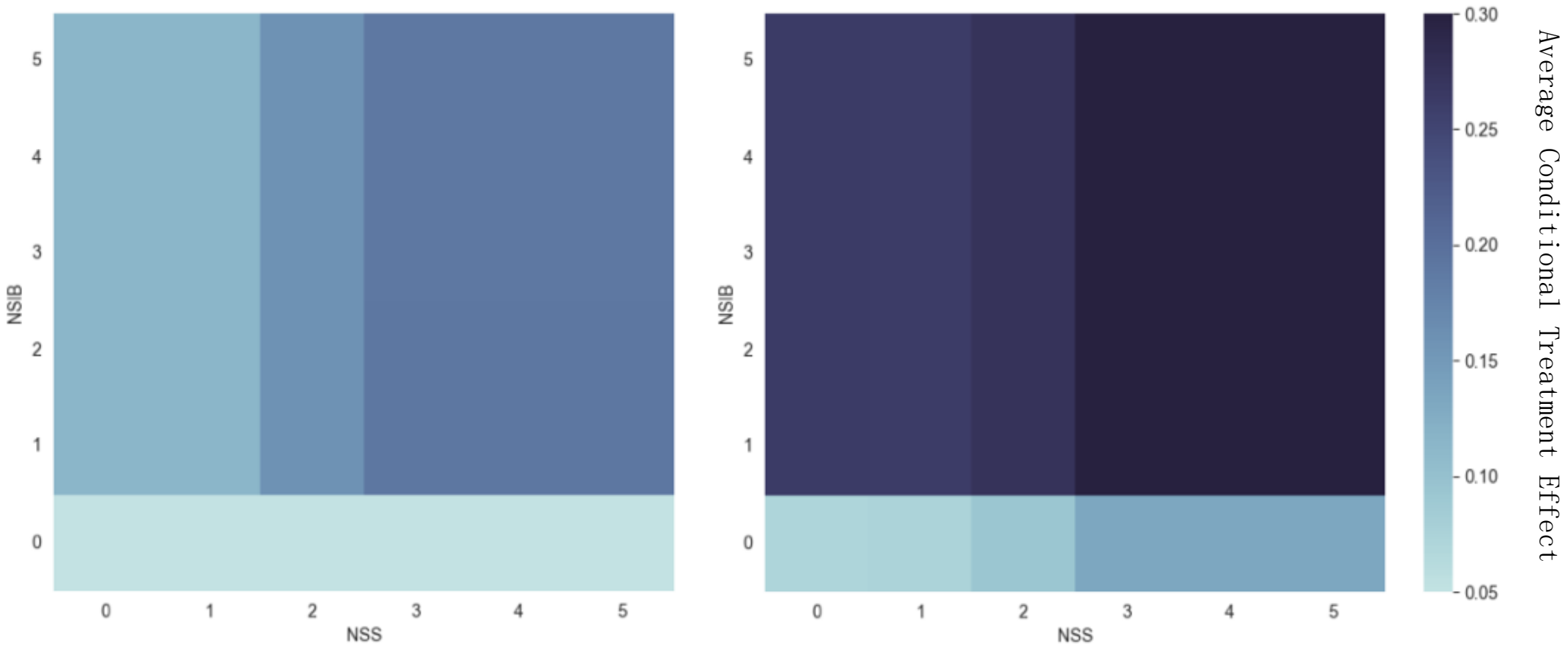

a) Conditioned on average level of survey response

b) Conditioned on median level of survey response

*Figure 4. The average treatment effect of walkability improvements, conditioned on two different levels of survey responses (average and median), as captured by NSIB (Number of scientific institutional buildings within 1 km) and NSS (Number of supermarkets or shopping malls within 1 km), The left panel displays the estimated treatment effect when conditioned on the average level of survey responses, while the right panel displays the effect when conditioned on the median level of survey responses. The horizontal and vertical axes represent the levels of NSIB and NSS, respectively. The color map represents the estimated treatment effect, with darker colors indicating a larger effect.*

# Discussion

This research employed survey data gathered from seven urban campuses in Wuhan city to ascertain the causal influence of campus walkability on the emotional aspects of walking experiences. After adjustment for various campus walkability attributes and individual attitude towards walking, we revealed a significant positive impact of improved campus walkability on affective walking experiences. Among various surveyed factors, the geographical features exerted the highest impact on students' affective walking experiences apart from the treatment of improved campus.

Counterfactual analysis was employed in this study as a valuable tool for estimating causal effects since experimental control was not feasible. It proved particularly useful in understanding the impact of the campus walkability intervention on the outcome of affective walking experiences. This analysis involved comparing the actual outcome with a hypothetical "counterfactual" outcome, which represented what would had occurred if the intervention had not been implemented. By conducting counterfactual analysis, more precise estimates of the causal effect can be obtained, especially when observed confounding factors were present, such as individual subjective attitudes towards walking experiences and surroundings.

In the context of our campus walkability study, the application of counterfactual analysis provided compelling evidence that changes in campus walkability indeed led to changes in affective walking experiences, rather than merely indicating an association between the two. This approach allows for a more rigorous evaluation of the causal relationship and helps avoid misleading conclusions that may arise from traditional statistical techniques like correlation and regression analyses, which could suggest that subjective factors are the primary determinants of the walking experience.

It was important to note that counterfactual analysis had its own assumptions and limitations, and it should be used in conjunction with other methods to ensure robust findings. The most important assumption was that of the 'potential outcomes' framework, which asserts that every individual has a well-defined potential outcome under each treatment level. This was an untestable assumption but necessary for the counterfactual framework to be valid. Another major assumption was the ignitability of treatment assignment, which may not hold if there were unobserved confounders that influence both the treatment assignment (i.e., the renovation of the campuses) and the outcome (the affective walking experience). As a result, findings from counterfactual analysis should be interpreted with caution.

Our study also had some limitations regarding data collection. Firstly, the survey data was gathered from only seven urban campuses in Wuhan, which may restrict the generalizability of the findings. Conducting similar research in diverse locations across China with varying cultural and regional backgrounds would greatly enhance our understanding of these relationships.

Secondly, the data primarily relied on survey responses, which lacked factual aspects. It would be valuable to incorporate factual features such as longitudinal spatial data, density variables, and land use mix to gain insights into changes in the physical environment over time and support more robust causal inference.

Furthermore, recent studies indicated that the characteristics of the built environment were associated with individuals' attitudes, suggesting that the walkable features of a campus could influence students' walking attitudes. Exploring the mediating role of walking attitude in the causal relationship between campus walkability and affective walking experiences in future studies would be intriguing.

By addressing these limitations, future research would provide a more comprehensive understanding of the relationships between campus walkability, affective walking experiences, and the underlying mechanisms that contribute to these associations.

# Conclusions

The use of counterfactual analysis provided compelling evidence for a causal relationship between the improvement in campus walkability and affective walking experiences, correcting the findings from conventional regression techniques. Geographical features were found to exert the highest impact on students' affective walking experiences, apart from the direct influence of campus improvements.

However, it was important to recognize the assumptions and limitations of counterfactual analysis, including the potential outcomes framework and the ignitability of treatment assignment. These findings would be interpreted cautiously and complemented with other research methodologies for robustness.

Further studies are necessary to deepen our understanding of these relationships and expand the scope of research beyond our current study. This would enable us to gain insights into mediating factors and cultural variations that influence the mental health of students on different campuses across China.

# Declaration

- **Ethics approval and consent to participate**

  The Ethics Committee of the Wuhan University granted approval for our survey. Written informed consent was obtained from all persons participating in the study/ their legal guardian(s) of all subjects, and the Ethics Committee of Wuhan University approved this consent procedure. This study was performed in accordance with the Declaration of Helsinki.

- **Consent for publication**

  Not applicable.

- **Availability of data and materials**

  The datasets used and/or analyzed during the current study is available from the corresponding author on reasonable request.

- **Competing interests**

  There are no conflicting interests with any individual or organization.

- **Funding**

  The present study is supported by the Fundamental Research Funds for the Central Universities, with grant number 20720221045.

- **Authors' contributions**

  Conceptualization, B.L. and J.Z.; methodology, B.L. and J.Z.; software, J.Z..; validation, B.L. and J.Z.; formal analysis, J.Z.; investigation, B.L.; resources, B.L.; data curation, B.L.; writing—original draft preparation, B.L. and J.Z.; writing—review and editing, B.L. and J.Z..; visualization, B.L. and J.Z.; supervision, B.L. and J.Z.; funding acquisition, B.L. All authors have read and agreed to the published version of the manuscript.

- **Acknowledgements**

  - Not applicable.